\documentclass [12pt,a4paper]{article}
\usepackage{graphicx}
\usepackage{cite}
\newcommand{\be}{\begin{equation}} 	\newcommand{\ee}{\end{equation}}
\newcommand{\ba}{\begin{eqnarray}} 	\newcommand{\ea}{\end{eqnarray}}
\newcommand{\bes}{\begin{equation*}}	\newcommand{\ees}{\end{equation*}}
\newcommand{\bas}{\begin{eqnarray*}}	\newcommand{\eas}{\end{eqnarray*}}
		\newcommand{\mc}{\mathcal}

\begin{document}
\centerline{\bf\Large Dark Energy and Dark Matter in 
General Relativity}
\centerline{\bf\Large with local scale invariance}

\bigskip
\centerline{\bf \large Pavan Kumar Aluri, Pankaj Jain and Naveen K. Singh}

\bigskip
\centerline{Physics Department, I.I.T. Kanpur, India 208016}
\centerline{emails: aluri@iitk.ac.in, pkjain@iitk.ac.in, naveenks@iitk.ac.in}

\bigskip
\begin{center}
\begin{minipage}{0.9\linewidth}
{\small {\bf Abstract:}
We consider a generalization of Einstein's general theory of relativity such
that it respects local scale invariance. This requires the introduction of
a scalar and a vector field in the action. We show that the theory 
naturally displays both dark energy and dark matter. We solve the resulting
equations of motion assuming an FRW metric. The solutions are found to be
almost identical
to those corresponding to the standard $\Lambda$CDM model.
}\end{minipage}\end{center}

\section{Introduction}
In recent papers \cite{JM,JMS} we have studied a scale invariant extension of the general
theory of relativity. The theory postulates a scalar field $\Phi$ and a
modification of the gravitational action such that \cite{Cheng,Cheng1},
\begin{equation}
{1\over 2\pi G} R \rightarrow \beta \Phi^*\Phi R
\end{equation}
where $G$ is the Newton's gravitational constant, $R$ is the Ricci scalar and
 $\beta$ is a dimensionless constant.
The modified action involves no mass parameter and is invariant under the
scale transformation. 
It is also convenient to introduce the concept of pseudo-scale
invariance. In four dimensions
the pseudo-scale transformation can be written as \cite{Cheng,Cheng1},
\ba
x & \rightarrow & x\, , \cr
\Phi &\rightarrow & \Phi/\Lambda\, ,\cr
g^{\mu\nu} &\rightarrow & g^{\mu\nu}/\Lambda^2\, , \cr
A_\mu &\rightarrow & A_\mu\, ,\cr
\Psi &\rightarrow & \Psi/\Lambda^{3/2}\, .
\label{eq:PSU}
\ea
where $x$ is the space-time coordinate, $g^{\mu\nu}$ the metric, $A_\mu$ a 
vector field and $\Psi$ a spin half field. The scale transformation can
be expressed as a combination of the pseudo-scale and the
general coordinate transformations. Hence as long as
general coordinate invariance is respected, pseudo-scale invariance is
equivalent to scale invariance.  

We also considered a theory with local scale or pseudo-scale invariance
\cite{Cheng,Cheng1,Padmanabhan85,Hochberg,Wood,Wheeler,Feoli,Pawlowski,Nishino,Demir}. 
The transformations in this case are given by Eq. (\ref{eq:PSU}) with 
the parameter $\Lambda\rightarrow \Lambda(x)$. 
In this case we need to introduce the Weyl vector meson, $S_\mu$,
besides the scalar field. 
Under pseudo-scale transformation the vector field transforms as 
\cite{Cheng,Cheng1}
\ba
S_\mu \rightarrow S_\mu - \frac 1f\partial_\mu ln(\Lambda(x))\,.
\ea
The Lagrangian in this case, ignoring all other fields besides the $\Phi$ and
$S_\mu$, may be written as
\ba
\mc L = -{\beta\over 8} \Phi^2 \tilde R+ \mc L_{matter}
\label{eq:Lagrangian}
\ea
where
\ba
\mc L_{matter} = \frac12g^{\mu\nu}(\mc D_\mu\Phi)(\mc D_\nu\Phi) -
{\lambda\over 4}\Phi^4 -
{1\over 4}g^{\mu\rho}g^{\nu\sigma}E_{\mu\nu}E_{\rho\sigma}\ ,
\ea
$f$ is the gauge coupling constant, $E_{\mu\nu}=\partial_\mu S_\nu-
\partial_\nu S_\mu$, $\tilde R$ is the modified curvature scalar
\cite{Cheng,Cheng1},
 invariant under local pseudo-scale transformation, and
 \ba
 \mc D_\mu &\equiv& \partial_\mu - fS_\mu
 \ea
 is the gauge covariant derivative.
The scalar $\tilde R$ is found to be
\ba
\tilde R = R - 6f S^\kappa_{\ ;\kappa} - 6f^2 S_\mu S^\mu
\ea
where $R$ is the standard curvature scalar.
The mass of the vector boson $S_\mu$ has been constrained by cosmological 
observations to be very light, less than 400 eV, or very heavy, greater than
Planck Mass \cite{Huang}.

A theory with local scale invariance is aesthetically pleasing since it 
does not contain any mass parameter in the action. This also implies that
we cannot add the cosmological constant term in the action. Hence as long 
as scale invariance is unbroken, cosmological constant is identically zero.
In Ref. \cite{JM,JMS} we argued that scale invariance is broken by the 
cosmological time evolution or equivalently by initial conditions. This
phenomenon is called cosmological symmetry breaking in Ref. \cite{JM,JMS}. 
The authors argue that the universe is a time dependent solution of
the equations of motion. All phenomena are described by making a quantum
expansion around this time dependent solution. Hence if the symmetries 
of the action are not respected by this solution then these symmetries 
will be hidden, in analogy to the phenomenon of spontaneous symmetry breaking. 
However as shown in Ref. \cite{JM,JMS}, cosmological symmetry breaking
is not the same as spontaneous symmetry breaking.

Once scale invariance is broken cosmologically, the theory generates
the dimensionful parameters such as the Newton's constant and the
vacuum (or dark) energy \cite{Wetterich,Ratra,Fujii,Chiba,Carroll1,Caldwell,Uzan,Amendola,Bean,Jain}. These essentially get related to the initial
conditions imposed on the scalar field. The fact that this theory leads
to dark energy has also been noticed in Ref. \cite{Wei}. However the precise
nature of identification is different from that in Ref. \cite{JM,JMS}. 
In Ref. \cite{JM,JMS} the authors speculated that if scale invariance 
is broken cosmologically, then this symmetry may not be anomalous. This 
symmetry may also tame the quantum corrections to the vacuum energy, hence
avoiding the fine-tuning problems in cosmological constant \cite{Weinberg,Peebles,Padmanabhan,Copeland,Carroll,Sahni,Ellis}.
Alternate proposals to solve the
cosmological constant problem are discussed in Ref.
\cite{Weinberg,Aurilia,VanDer,Henneaux,Brown,Buchmuller,Henneaux89,Sorkin,Sundrum}.

In the case of global scale invariance, the authors in Ref. \cite{JMS} found
a solution with $\Phi$ equal to a constant. 
In this case our theory reduces to a scalar-tensor model. The 
cosmological implications of such models
have been studied extensively in recent literature
\cite{Agarwal,Barrow,Barenboim}.
The value of the constant field $\Phi$ can
be chosen to fit the experimental value of the gravitational constant.
For local scale invariance, $\Phi$ is constant in the gauge $S_0=0$. 
In Ref. \cite{JMS}, the authors set $S_i=0$, where the index $i=1,2,3$. 
However in general $S_i$ is not zero. In the present paper we give a general
solution to the equations of motion of this model. We find that the solutions
naturally lead to both dark energy and dark matter. 

In Ref. \cite{JM,JMS}, the authors had suggested that the scalar field $\Phi$ 
may be the Higgs multiplet. They argued that cosmological symmetry breaking
also leads to a breakdown of the electroweak symmetry. In this theory
the Higgs particle is absent from the particle spectrum and acts as the
longitudinal mode of the vector field $S_\mu$ \cite{Cheng,Cheng1}. 
Although this is
 a consistent picture, the field $\Phi$ may also be associated with a
 scalar field in a grand unified theory (GUT).
 In the present paper, we assume $\Phi$ to be a real
 scalar field. The generalization to the case where $\Phi$ may be a standard
 model or a GUT scalar field multiplet is straightforward.

\section{Equations of motion}
The Einstein's equations and the equations of motion for $\Phi$ and $S_\mu$
following from Eq. (\ref{eq:Lagrangian}) are given in Ref. \cite{JMS}. Here
we assume that, at leading order,
all the fields are independent of space coordinate and 
depend only on time. In this case the equations simplify considerably. 
We display these equations below, correcting some typographical errors
in Ref. \cite{JMS}.
The time-time and space-space components of the Einstein's equation are
\ba
\Phi^2\left(R_{00} -{R\over 2}\right) + 3\Phi^2f^2(S^iS_i - S_0S_0)
+ 3f S_0 \partial_0(\Phi^2) - g^{ij}\partial_0(\Phi^2) \Gamma^0_{ij} =
{4\over \beta} T_{00}\, ,
\ea
and
\ba
\Phi^2\left(R_{ij} - {1\over 2} g_{ij} R\right)
+ 3f^2\Phi^2 g_{ij} S_\mu S^\mu &-& 6f^2\Phi^2 S_iS_j
- 3fg_{ij} S^0\partial_0\Phi^2\nonumber\\
&+& g_{ij} \partial_0\partial_0\Phi^2
-2a\dot a \delta_{ij}\partial_0\Phi^2
= {4\over\beta} T_{ij}
\ea
respectively. Here dots represent derivatives with respect to time.
The equation of motion for the scalar field is
\ba
\ddot \Phi - f\Phi \dot S_0 + 3\dot \Phi{\dot a\over a} - f^2S_\mu S^\mu\Phi
+\lambda\Phi^3 - 3f\Phi S_0{\dot a\over a} + {\beta\over 4} \Phi\tilde R = 0
\ea
The corresponding equations for $S_0$ and $S_i$ are
\ba
f S_0 = {\dot\Phi\over \Phi}\,,
\label{eq:S0}
\ea
\ba
\ddot S_i + {\dot a\over a}\dot S_i + f^2\Phi^2 S_i + {3\over 2}\beta f^2\Phi^2
S_i = 0
\label{eq:Si}
\ea
respectively.
Since the theory has local scale invariance, we need to fix the gauge in order
to obtain a unique solution. We choose the gauge $S_0=0$. In this case 
Eq. (\ref{eq:S0}) implies that the scalar field $\Phi$ is a constant, 
independent of time. 

Thus, we set $\Phi(t) = \eta={\rm constant}$ with the above mentioned gauge choice. The 
nonzero constant value of $\eta$ essentially acts as dark energy 
in the present model. The existence of dark energy is naturally implied
by cosmologically broken scale invariance.
The resulting 0-0 and i-j components of the 
Einstein's equations can then be written as, 
\begin{eqnarray}
3\eta^2 {\dot{a}^{2} \over a^{2}}={4 \over \beta} \left[{\lambda \over 4} \eta^{4} + {\dot{S}_i^{2} \over 2a^{2}}+\left(1+{3\beta \over 2}\right) 
{f^{2}\eta^{2}S_i^{2} \over 2a^{2}}\right]
\label{eqMEE1}
\end{eqnarray}
\\
and
\begin{eqnarray}
3\eta^{2}\left(2{\ddot{a} \over a}+{\dot{a}^{2} \over a^{2}}\right) = 
{4 \over \beta} \left[{3\lambda \over 4}\eta^{\,4}-{\dot{S}_i^{\,2} \over 2a^{\,2}}+\left(1+{3\beta \over 2}\right){f^{\,2}\eta^{\,2}S_i^{\,2} \over 2a^{\,2}}\right]
\label{eqMEE2}
\end{eqnarray}\\
respectively.
In Eqs. \ref{eqMEE1} and \ref{eqMEE2}, sum over the subscript $i$ is implied
in terms containing $S_i^2$ or $\dot S_i^2$.
The equations for $\eta$ and $S_i$ become,
\begin{equation}
3\eta^2\left({\ddot{a} \over a}+{\dot{a}^2 \over a^2}\right)={4 \over \beta}\left[{\lambda \over 2}\eta^2+\left(1+{3\beta \over 2}\right) \frac{f^2S_i^2\eta^2}{2a^2}\right]\,,
\label{eqEta}
\end{equation} 
\begin{equation}
\ddot{S}_i+{\dot{a} \over a}\dot{S}_i+\left(1+{3\beta \over 2}\right)f^{\,2}\eta^{\,2}S_i =0\,
\label{eqSi}
\end{equation}
respectively.

The background constant value of the field $\Phi$ can be related to the 
Planck mass, $M_p$, by the relation \cite{Cheng,Cheng1,JM,JMS},
\begin{equation}
\beta\eta^2 = {M_p^2\over 2 \pi}
\end{equation}
The mass of the vector field, $S_\mu$, is found to be,
\begin{equation}
M_S^2 = \left(1+{3\beta\over 2}\right)f^2\eta^2\,. 
\end{equation}
If the vector field $S_\mu=0$, then
the Hubble constant, $H_0$, 
is given by, 
\begin{equation}
H_0 = \sqrt{{\lambda\over 3\beta}}\,\eta = \sqrt{\lambda\over 6 \pi}\, {M_p\over \beta}\,, 
\label{eq:hubble1}
\end{equation}
Finally the vacuum energy density is given by,
\begin{equation}
\rho_\Lambda = {1\over 4}\lambda \eta^4. 
\label{eq:vac}
\end{equation}
From the relationship between the Hubble constant and the Planck mass, it is clear that either $\lambda$ is extremely small or $\beta$ is extremely
large. If we assume $\beta \sim O(1)$, then $\lambda$ is found to be of
order $10^{-60}$. In Ref. \cite{JM,JMS} the authors argued that this small
value by itself may not lead to fine tuning problems. These would arise if
the quantum corrections at each order require very delicate cancellations
to maintain the small value of this parameter. The quantum corrections 
in this model have so far not been computed. It is of course also important
to explain the origin of such a small parameter. This issue is beyond the scope
of the present paper. However we speculate that such small values of scalar
coupling might arise due to the well known triviality of scalar field theories
\cite{Aizenman,Frohlich}.
The continuum scalar field self coupling is driven to zero by 
renormalization group analysis. However at length scales smaller than the 
Planck length, it may not be appropriate to treat space-time as a continuum. 
The discrete nature of space time might generate a small value for the
scalar self coupling. 

\section{Cosmological  Solution}
The equation of motion for $S_i$, Eq. (\ref{eqSi}), is similar to that of a damped harmonic oscillator,
with weakly time-dependent frequency and decay terms. 
We seek a solution of the form
\begin{equation}
S_i = n_i\, {\cal S}
\label{eq:ni}
\end{equation}
where $n_i$ is a constant unit vector.
The solution for ${\cal S}$ can be expressed as, 
{\large
\begin{eqnarray}
{\cal S} &=& Re\left\{Ae^{-\int\!{\dot{a} \over 2a}dt\,-\,i\int\!\omega_1dt}+
Be^{-\int\!{\dot{a} \over 2a}dt\,+\,i\int\!\omega_1dt}\right\}\,,
\label{eqSisol}
\end{eqnarray}}
where, \(A\) and \(B\) are assumed to be slowly varying functions of time and \(\omega_1^2=\omega^2-{H^2 \over 4}\, ,\, \omega^2=\left(1+{3\beta \over 2}\right)
f^{\,2}\eta^{\,2}=M_S^2\,\, \textrm{and} \,\, H = \dot{a}/a \).
By substituting this solution in Eq. (\ref{eqSi}) and neglecting second derivatives of \(A\) and \(B\), we get,
{\large
\begin{eqnarray}
\begin{array}{l l l}
{\dot{A} \over A} + {\dot{\omega}_1 \over 2\,\omega_1}=i{\dot{H} \over 4\omega_1} 
& \Rightarrow & A={k_1 \over \sqrt{\omega_1}}e^{{i \over 2}sin^{-1}{H \over 2\omega}}\\
{\dot{B} \over B} + {\dot{\omega}_1 \over 2\,\omega_1}=-i{\dot{H} \over 4\omega_1} 
& \Rightarrow & B={k_2 \over \sqrt{\omega_1}}e^{-{i \over 2}sin^{-1}{H \over 2\omega}}
\end{array}
\label{ABsol}
\end{eqnarray}}
where, $k_1$ and $k_2$ are constants of integration and are, in general, complex.
Since $\omega>>H$ we see that $A$ and $B$ vary very slowly with time 
compared to other terms in $S_i$. The most rapidly varying terms are those containing
$\int\omega_1 dt$ in the exponent. Due to these terms, $S_i$ fluctuates rapidly
with time. 

Due to the presence of a vector field $S_i$ in our theory, our cosmological
solution naturally contains a constant three dimensional unit vector 
$n_i$ defined in Eq. \ref{eq:ni}. This vector defines a direction in space
and hence breaks rotational invariance. However the background metric is
still isotropic since the vector $n_i$ does not contribute to Einstein's
equations. Furthermore it is unlikely to lead to very large observable 
consequences of the breakdown of isotropy. This is because the field $S_\mu$
does not directly interact with visible matter \cite{Cheng,Cheng1}. 
Nevertheless, it is extremely interesting to determine the cosmological 
predictions of this breakdown of isotropy in view of several observations
which indicate a preferred direction in the universe 
\cite{Birch,JP99,huts1,huts2,Oliveira,Ralston,Schwarz,Eriksen}.
Models in which vector fields acquire nonzero vacuum or background 
values have also
been considered by many authors \cite{Ford,Lidsey,Armendariz,Alves,Benini,Wei1,Ackerman,Dulaney1,Yokoyama,Golovnev,Kanno,Koivisto,Bamba,Chiba1,Dulaney,Dimopoulos}. 
It has been argued that many of these models, which lead to prolonged
anisotropic accelerated expansion, are unstable \cite{Himmetoglu}.
In our model the vector field does not directly lead to anisotropic 
expansion, even though it acquires a non-zero background value.

\subsection{Leading Order Solution}
At leading order we can assume that $A$ and $B$ are time independent. The
leading order solution for ${\cal S}$ can then be written as
\begin{equation}
{\cal S} = {1\over \sqrt{a}} (A' \cos\theta + B'\sin\theta)
\end{equation}
where $\theta = \int \omega_1 dt$ and $A'$ \& $B'$ are some real constants.
We also find
\begin{equation}
\dot {\cal S} = {1\over \sqrt{a}} \left(-{H\over 2}p + \omega_1 q\right)\, ,
\end{equation}
where, $p = A'\cos\theta + B'\sin\theta$, $q=-A'\sin\theta + B'\cos\theta$.
We next substitute these into the 0-0 component of the Einstein's equation. 
We define,
\begin{equation}
\rho_{S_i} = {1\over 2a^2}\dot S_i^2 + {1\over 2a^2}\omega^2 S_i^2
\label{eq:rhoSi}
\end{equation}
This essentially acts as the contribution to the energy density provided
by the field $S_i$. We find,
\begin{equation}
\rho_{S_i} = {1\over 2a^3}\left[\omega^2(p^2+q^2) + {H^2\over 4}
(p^2-q^2) - \omega_1 H pq\right] 
\end{equation}
where, $p^2+q^2 = A'^2+B'^2$, which is a constant. Since $\omega_1$ is very large,
$S_i$ is a rapidly oscillating function of time. Hence it is reasonable to
replace the oscillatory functions with their time averages. After averaging
over time, $(p^2-q^2)\rightarrow 0$ and $pq\rightarrow 0$. Hence,
\begin{equation}
\rho_{S_i} = {1\over 2a^3}\left(A'^2 + B'^2\right)\omega^2
\end{equation}
We next consider the i-j component of the Einstein's equation. 
We define,
\begin{equation}
-3P_{S_i} = -{1\over 2a^2}\dot S_i^2 + {1\over 2a^2}\omega^2 S_i^2
\label{eq:PSi}
\end{equation}
which effectively acts as the contribution of the $S_i$ field to pressure.
After substituting the time averaged values for the oscillatory functions, we get
\begin{equation}
P_{S_i} = 0 
\end{equation}
Hence, we find that the field $S_i$ essentially acts as the cold dark matter. 
Its energy density $\rho_{S_i}$ varies as $1/a^3$ and its pressure 
$P_{S_i}$ is zero at leading order. A similar phenomenon is seen in the
case of coherent axion oscillations 
\cite{Preskill,Abbott,Dine,Steinhardt,KT}.

The modified Einstein's equations, at leading order, can now be written as
\begin{eqnarray}
H^2 = {\dot a^2\over a^2} &=& {\lambda\over 3\beta}\eta^2 + {2\over 3\beta\eta^2 a^3}
(A'^2+B'^2)\omega^2
\label{eq:H} \\
2{\ddot a\over a} + {\dot a^2\over a^2} &=& {\lambda\over \beta}\eta^2\,.
\label{eq:ddota}
\end{eqnarray}
Eq. (\ref{eq:H}) generalizes the expression for the Hubble constant,
Eq. (\ref{eq:hubble1}), for the case when the vector field is non-zero.

\subsection{Corrections to the leading order}
We next calculate the corrections to the leading order result by taking
into account the time dependence of the coefficients $A$ and $B$ in Eq. (\ref{eqSisol}). Substituting for $A$ and $B$, from Eq. (\ref{ABsol}), in Eq. (\ref{eqSisol}), 
we find,
\begin{eqnarray}
{\cal S} &=& {1 \over \sqrt{\omega_1 a}}\left[Q\cos(\theta-x) + P\sin(\theta-x)
\right]={1\over \sqrt{\omega_1 a}} U\\
\dot {\cal S} &=& {1 \over \sqrt{\omega_1 a}}\left[{H\over 2}\left({\dot x\over 
\omega_1} -1\right)U + (\omega_1-\dot x)V\right]
\end{eqnarray}
where, 
\begin{eqnarray*}
U &=& N\cos\theta + M\sin\theta\ , V = -N\sin\theta+M\cos\theta \\
\textnormal{and}\,\,\,M &=& P\cos x + Q\sin x\ , N = -P\sin x+Q\cos x\,.
\end{eqnarray*}
Here, $x = {1 \over 2}\sin^{-1}{H \over 2\omega} = {1 \over 2}\cos^{-1}{\omega_1\over
\omega}$, $\dot x = \dot H/4\omega_1 = -\dot\omega_1/H$ and $P$ \& $Q$ 
are some real constants.

Substituting these in Eq. (\ref{eq:rhoSi}), we get,
\begin{eqnarray}
\rho_{S_i} = {1\over 2\omega_1a^3}\Bigg[(U^2+V^2)\omega^2
&+&{H^2\over 4}(U^2-V^2) - \omega_1H\left(1-{\dot x\over \omega_1}\right)^2UV
\nonumber\\
&+& (\dot x^2-2\omega_1\dot x)\left({H^2\over 4\omega_1^2}U^2 + V^2\right)
\Bigg]\, . 
\end{eqnarray}
The third term on the right hand side reduces to $\omega_1HUV$, since $\dot x/
\omega_1<<1$. The fourth term simplifies to $-\dot H V^2/2\,$, if we
neglect terms suppressed by factors of $H/2\omega_1$. Hence we find
\begin{equation}
\rho_{S_i} = {1\over 2\omega_1a^3}\Bigg[(U^2+V^2)\omega^2
+{H^2\over 4}(U^2-V^2) - \omega_1HUV
- {\dot H\over 2}V^2
\Bigg]\, . 
\end{equation}
We again substitute time averaged values for rapidly oscillating functions. 
This sets $(U^2-V^2)\rightarrow 0$, $UV\rightarrow 0$ and $(U^2+V^2)
=(M^2+N^2)= P^2+Q^2$, which is a constant. A leading order expression for
$\dot H$ can be computed using Eq. (\ref{eq:H}). We find
\begin{equation}
\dot H = -{1\over \beta\eta^2a^3}(A'^2+B'^2)\omega^2\,.
\end{equation}
Thus, we get,
\begin{equation}
\rho_{S_i} = {(P^2+Q^2)M_S\over 2 a^3} + {(P^2+Q^2)\lambda\eta^2
\over 48\beta a^3 M_S} 
+{(P^2+Q^2)(A'^2+B'^2)M_S \over 6\beta\eta^2 a^6}\,. 
\end{equation}
The leading term varies as $a^{-3}$ as already found in the previous
section. Here, we also find two subleading terms. One of these falls as
$1/a^3$ and the second falls much faster, as
$a^{-6}$, as the universe expands. 
We similarly find the corrections to the pressure term $P_{S_i}$. We find that,
using Eq. (\ref{eq:PSi}),
\begin{equation}
-3P_{S_i} = {1\over 2a^3\omega_1}\Bigg[(U^2-V^2)\omega_1^2
+\omega_1HUV + {\dot H\over 2}V^2 \Bigg]\, . 
\end{equation}
Again, substituting time averaged values for the rapidly oscillating functions,
we get
\begin{equation}
P_{S_i} = -{1 \over {6\,\omega_1 a^3}}{\dot{H} \over 4}(P^2+Q^2) 
= {(P^2+Q^2)(A'^2+B'^2)M_S \over 24\beta\eta^2 a^6}\,.
\end{equation}
Hence, we get a small correction term to $P_{S_i}$, which also decays rapidly
as $a^{-6}$ as the universe expands.

The 0-0 and i-j component of the Einstein's equations can, now, be written as, 
\begin{equation}
{3\beta \over 4}\eta^2H^2 = {\lambda \over 4}\eta^4 + {(P^2+Q^2) \over 2\omega_1a^3} \left(\omega^2-{\dot{H} \over 4}\right)
\label{eqMEE1sim}
\end{equation}
and
\begin{equation}
{3\beta \over 4}\eta^2\left(2{\ddot{a} \over a} + {\dot{a}^2 \over a^2}\right)= 
{3\lambda \over 4}\eta^4 + {(P^2+Q^2) \over 2\omega_1 a^3}{\dot{H} \over 4}
\label{eqMEE2sim}
\end{equation}
respectively. The first of the above two equations can be cast in the form,
\begin{equation}
1 = \Omega_\Lambda + \Omega_{S_i}
\label{eqSumComp}
\end{equation}
where
\[
\Omega_\Lambda = {\rho_\Lambda \over \rho_{cr}}\,,\Omega_{S_i} = {\rho_{S_i} \over \rho_{cr}} \,,
\]
\begin{eqnarray*}
\rho_\Lambda = {\lambda \over 4}\eta^4\,\,,  
\rho_{S_i} = {(P^2+Q^2) \over 2\omega_1a^3}\left(\omega^2-{\dot{H} \over 4}\right)\,\, {\rm and} \,\, \rho_{cr} = {3\beta \over 4}\eta^2H^2\,.
\end{eqnarray*}
Eq. (\ref{eqSumComp}) looks like the $\Lambda$CDM model with $\Omega_M = \Omega_{S_i}$.

Thus, the energy density $\rho_{S_i}$ and the corresponding pressure $P_{S_i}$ of the vector field $S_i$, including the correction terms, are obtained as
\[
\rho_{S_i} = {c_1 \over 2\omega_1a^3}\left(\omega^2 + {c_2 \over a^3}\right)
\]
and
\[
P_{S_i}= {c_1c_2 \over 6\,\omega_1a^6}
\]
where, 
$$ c_1 = P^2+Q^2\ ,\ c_2={A'^2+B'^2\over 4\beta}\left(1+{3\beta\over 2}\right)f^2\ .$$
In the limit $x\rightarrow 0$ and $\omega_1\rightarrow \omega$, we find
$A'=Q/\sqrt{\omega}$ and $B'=P/\sqrt{\omega}$. 

We can make an estimate of the term $(P^2+Q^2)$. Since the recent cosmological observations support a flat $\Lambda$CDM model, we can equate the second term on right hand side of Eq. (\ref{eqMEE1sim}), evaluated at present time, to $\rho_{M,0}$. The contribution due to $\dot{H}$ is negligible. Thus, we get,
\[
P^2+Q^2 \approx {3M_P^2 H_0^2\Omega_{M} \over 4 \pi M_S }
\]
where $\Omega_M$ is computed at the current time.
\section{Including the contribution due to radiation}
In this section, we obtain a set of dynamical equations to study the evolution of different components of the universe since the beginning of the 
radiation dominated era. For this purpose we introduce, by hand, the contribution due to radiation. We expect to reproduce the usual Big Bang evolution where radiation dominates at early times, followed by dark matter and dark energy dominated eras, respectively, at late times. We introduce a radiation term with energy density $\rho_R $ and it's corresponding pressure term $P_R $ in the energy-momentum tensor $T_{\mu \nu}$. The resulting equations are solved numerically.

It is convenient to introduce the following variables,
\begin{eqnarray}
&& X^2 = \frac{\lambda}{3\beta}{\eta^2 \over H^2} = \Omega_\Lambda\,, \\ \nonumber
&& Y^2 = {2 \over 3\beta}{{\cal S}^2 \over a^2\eta^2} = \Omega_1\,, \\ \nonumber
&& Z^2 = {2 \over 3\beta}\left(1+{3\beta \over 2}\right){{f^2 {\cal S}^2} \over {a^2 H^2}} = \Omega_2\,, \\
&& R = \frac{4}{3\beta}{\rho_{R,0} \over a^4\eta^2H^2}=\Omega_R \,. \nonumber
\end{eqnarray}
Here, $\Omega_{S_i} = \Omega_1 + \Omega_2$, $\rho_{R,0}$ is the 
radiation energy density in the current era and 
the prime denotes derivative with respect to $\ln a$. 
Hence for any function $f$,
$$f' \equiv {df\over d\ln a}={1\over H}{df\over dt}\,.$$
With these variables, we can cast the equations (\ref{eqMEE1}), (\ref{eqEta}) and (\ref{eqSi}), along with $\rho_R$ and $P_R$, in a dimensionless form, to obtain the following set of equations :
\begin{eqnarray}
&& X' = X(2-2X^2-Z^2) \,, \\\nonumber
&& Y' = -Y(2X^2+Z^2)-{\kappa \over 2}XZ \,, \\ \nonumber
&& Z' = Z(1-2X^2-Z^2)+{\kappa \over 2}XY \,, \\ \nonumber
&& R' = -2R(2X^2+Z^2) \,,
\end{eqnarray}
where, $\kappa =\sqrt{{12\beta \over \lambda}}{\omega \over \eta} = 
 \sqrt{3}M_PM_S/\sqrt{2\pi\rho_V}   $.

\begin{center}
\begin{figure}
\includegraphics[angle=-90,width = 0.94\textwidth]{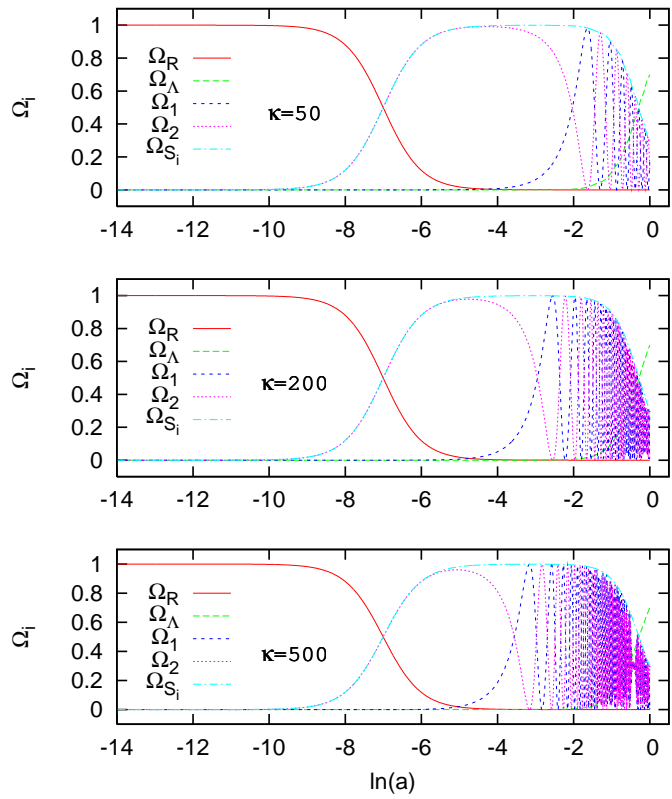}
\caption{The ratio of energy density to the critical energy density, 
$\Omega_i$, for different components as
 a function of $\ln(a)$ for $\kappa=50,200, 500$.}
\label{omega-i}
\end{figure}
\end{center}

We studied the dynamical equations numerically from the beginning of 
radiation dominated era ($\ln a = -29$) till today ($\ln a = 0$).
The results are presented in Fig. (\ref{omega-i}). In the graphs we only show
results for the range $\ln a = [-14,0]$, as radiation is the only dominant 
component in the omitted regions. The plots show the results for 
three values of $\kappa=50, 200, 500$. The initial conditions for these
three cases have been chosen so as to match the final observed values
of $\Omega_M$ and $\Omega_\Lambda$ \cite{Dodelson,Essence,Kowalski,wmap}. 

As is evident from the plots, varying $\kappa$ varies the frequency of oscillations. Besides that the results are almost identical, as long as $\kappa>>1$. 
This can be understood from the expression of $\kappa$. 
For fixed values of $M_P$ and $\rho_V$, 
increasing $\kappa$ increases $\omega$ or
$M_S$, which implies more rapid oscillations. Furthermore as seen
from our analytic results, applicable when radiation energy density
is negligible, we reproduce the standard $\Lambda CDM$ model in the large
$\kappa$ limit.

\section{Conclusions}
We have analyzed a locally scale invariant generalization of Einstein's 
gravity. The theory requires introduction of a scalar and a vector 
field. The scale invariance in the theory is broken by a recently introduced
mechanism called the cosmological symmetry breaking. 
We have shown that this theory naturally leads to both dark energy and 
dark matter. Due to scale invariance the cosmological constant term is
absent in the action. The solutions to the equations of motion admit a 
constant, non-zero value of the scalar field, which leads to a small 
cosmological constant or dark energy. The cold dark matter arises in the
form of vacuum oscillations of the vector field. 
We have shown that the theory behaves very similar to the $\Lambda CDM$ 
model with negligible corrections. The precise values of the energy densities
of different components are fixed by the initial conditions. Some of the
parameters in the model take very small values and it is necessary 
to find an explanation for such small values. Furthermore it is 
important to compute quantum corrections in this model since that
will determine whether the model suffers from fine tuning problems. 
The model can be generalized to include the standard
model fields. The 
scalar field may then be identified with the Higgs multiplet. In this
case the Higgs particle is absent from the particle spectrum and hence
provides a very interesting test of the model. Alternatively the scalar
field might be identified with a GUT scalar field multiplet. This possibility
has so far not been studied in the literature.

\bigskip

{\bf Acknowledgements :} Pavan Kumar Aluri and Naveen Kumar Singh thank 
the Council of Scientific and Industrial Research(CSIR), India for providing 
their Ph.D. fellowships. Their fellowship numbers are  
F.No.09/092(0413)/2005-EMR-I and F.No.09/092(0437)/2005-EMR-I, respectively.

\end{document}